\documentclass[aps,preprint,showpacs]{revtex4}
\usepackage{amsfonts}
\usepackage{amsmath}
\usepackage{amssymb}
\usepackage{graphicx}
\usepackage{epsfig}

\newcommand{\iu}{\mathbf{i}}

\begin{document}

\title{New integrals in few-body problems.}

\author{E.~Z. Liverts}
\affiliation{Racah Institute of Physics, The Hebrew University, Jerusalem 91904,
Israel}

\author{N. Barnea}
\affiliation{Racah Institute of Physics, The Hebrew University, Jerusalem 91904,
Israel}

\begin{abstract}
This work is concerned with multi-dimensional integrals, which are making their appearance
in few-body atomic and nuclear physics. It is shown that the relevant two- and three-dimensional
integrals can be reduced to one-dimensional form. This implies that the internal one- and two-dimensional
integrals can be evaluated in explicit analytic form in term of the familiar
generalized hypergeometric functions. Some of the integrals are presented here for the first time.
\end{abstract}


\maketitle

\section{Introduction}

One of the effective tools for solving, e.g., $N$-body problem both in atomic and 
nuclear physics is introducing the Jacobi vectors (coordinates) $\boldsymbol\xi_i$ which presents a linear
combination of a standard vectors $\textbf{r}_i$  ($i=1,2,..N$) of the particles under consideration.
Omitting the center of mass motion
for a given choice of the Jacobi vectors, the \textit{hyperspherical coordinates} are given by 
the so-called  \textit{hyperradius}  $\rho$ and by a set $\Omega_{N-1}$ of angular variables.
The latter can be expressed through the $2(N-1)$  polar angles  $\omega_i\equiv(\theta_i,\phi_i)$
of the Jacobi vectors $\boldsymbol\xi_i$ and $(N-2)$ \textit{hyperspherical angles} $\varphi_i (i=2,...,N-1)$. 

In order to estimate the physical properties of the considered system, one needs to calculate matrix elements of the proper Hamiltonian in the appropriate basis.
 
For simplicity let us consider three-particle system (see, e.g., \cite{KI,B1}) in the most abundant basis set presenting the product
\begin{equation}
\label{1}
R_n(\rho) \mathcal{Y}_{\{l_1,l_2,\mu\},LM}(\Omega_2),
\end{equation}
where 
\begin{equation}
\label{2}
\mathcal{Y}_{\{l_1,l_2,\mu\},LM}(\Omega_2)=\mathcal{P}^{l_1,l_2}_\mu(\varphi_2)
\sum_{m_1,m_2}\langle l_1m_1l_2m_2\arrowvert LM  \rangle Y_{l_1m_1}(\theta_1,\phi_1)Y_{l_2m_2}(\theta_2,\phi_2)
\end{equation}
is the so called \textit{hyperspherical harmonic} function with definite angular momentum associated to the quantum numbers $LM$. The RHS of Eq.(\ref{2}) contains the function
\begin{equation}
\label{3}
\mathcal{P}^{l_1,l_2}_\mu(\varphi_2)=\mathcal{N}^{l_1,l_2}_\mu(\cos \varphi_2)^{l_2}
(\sin \varphi_2)^{l_1}P_\mu^{l_1+1/2,l_2+1/2}(\cos 2\varphi_2),
\end{equation}
where $P_\mu^{a,b}(z)$ are the Jacobi polynomials, $\mathcal{N}^{l_1,l_2}_\mu$ is normalization
constant; the Clebsch-Gordan coefficients $\langle l_1m_1l_2m_2\arrowvert LM  \rangle$, and usual
spherical harmonics $ Y_{lm}(\theta,\phi)$. Here $\mu$ is non-negative integer, and $(l_i,m_i)$ are 
the quantum numbers corresponding to the angular momentum operator associated with the $i$th Jacobi vector.

The most commonly encountered form of hyperradial function \cite{KI,B1} is the following:
\begin{equation}
\label{4}
R_n(\rho)=\mathcal{C}_{\alpha ,n}L_n^{(5)}(\alpha \rho)e^{-\frac{1}{2}\alpha \rho},
\end{equation}
where $L_n^{(k)}(z)$ are the generalized Laguerre polynomials, and $\alpha$ is a scale factor.
The explicit forms of the normalization constants $\mathcal{C}_{\alpha ,n}$ and $\mathcal{N}^{l_1,l_2}_\mu$
are not important for the given treatment, in contrast to the form of Jacobi vectors
\begin{equation}
\label{5}
{\boldsymbol\xi}_1=\frac{1}{\sqrt{6}}\left( 2\bf{r}_3-\bf{r}_1-\bf{r}_2\right) ,
~~~{\boldsymbol\xi}_2 =\frac{1}{\sqrt{2}}\left( \bf{r}_2-\bf{r}_1\right) 
\end{equation}
for three particles of equal mass (see, e.g., \cite{KI,BN,ER}).
Taking the inverse transformation for Jacobi vectors (\ref{5}) plus the centre-of-mass vector
${\boldsymbol\xi}_3=\frac{1}{3}\left( \bf{r}_1+\bf{r}+\bf{r}_3\right) $ 
and keeping in mind that by definition ${\xi}_2=\rho \cos(\varphi_2), \xi_1=\rho \sin(\varphi_2) $ 
one obtains:
\begin{equation}
\label{6}
r^2_{12} = {\rho}^2 (1+\tau),
\end{equation}
\begin{equation}
\label{7}
r^2_{13} = \frac{\rho^2}{2}\left[2-\tau+\lambda\sqrt{3(1-\tau^2)} \right], 
\end{equation}
\begin{equation}
\label{8}
r^2_{23} = \frac{\rho^2}{2}\left[2-\tau-\lambda\sqrt{3(1-\tau^2)} \right], 
\end{equation}
where $\tau=\cos 2\varphi_2$, and $\lambda$ presents cosine of the angle between vectors 
${\boldsymbol\xi}_1$ and ${\boldsymbol\xi}_2$. 

Let us consider the Gaussian-type central potentials of the form
\begin{equation}
\label{9}
V(\underset{i>j}{\{r_{ij}\}})=\sum_{i>j}\sum_k \mathcal{A}_k \exp(-\zeta_k r_{ij}^2).
\end{equation}
The examples of such potentials can be presented by the two-nucleon potential models \cite{VO,AF,TH}.

It can be shown that calculations of the matrix elements for the potentials of the form (\ref{9}) reduce to evaluating two-dimensional integral of the form 
\begin{eqnarray}
\label{10}
\mathcal{I}_2=\int^1_{-1}(1-\tau)^{l_1+1/2}(1+\tau)^{l_2+1/2} P^{l_1+1/2,l_2+1/2}_{\mu_1}(\tau)P^{l_1+1/2,l_2+1/2}_{\mu_2}(\tau)d\tau \times
~~~~~~~~~~~\nonumber~\\
\int^\infty_0 L^{(5)}_{n_1}(x)L^{(5)}_{n_2}(x)e^{-\gamma x^2-x} x^5 dx, 
\end{eqnarray}
and three-dimensional integral of the form 
\begin{eqnarray}
\label{11}
\mathcal{I}_3=\int^1_{-1}(1-\tau^2)^{} 
P^{l_1+1/2,l_1+1/2}_{\mu_1}(\tau)P^{l_2+1/2,l_2+1/2}_{\mu_2}(\tau)d\tau \times
~~~~~~~~~~~~~~~~~~~~~~~~\nonumber~\\
\int^\infty_0 L^{(5)}_{n_1}(x)L^{(5)}_{n_2}(x)e^{-\beta x^2-x} x^5 dx
\int_{-1}^1e^{-\lambda \kappa x^2}P_{l_1}(\lambda)P_{l_2}(\lambda)d\lambda, 
\end{eqnarray}
where
\begin{equation}
\label{12}
\gamma\equiv\gamma(\tau)=\frac{\zeta_k}{\alpha^2}(1+\tau),
\end{equation}
\begin{equation}
\label{13}
\beta\equiv\beta(\tau)=\frac{\zeta_k}{\alpha^2}\left(1-\frac{\tau}{2} \right),~~~\kappa\equiv\kappa(\tau)=\frac{\zeta_k}{2\alpha^2}\sqrt{3(1-\tau^2)},
\end{equation}
and $P_l(\lambda)$ are the Legendre polynomials.
One should emphasize that the 3-dimensional integral (\ref{11}) corresponds only to the case of
zero angular momentum ($L=0$). However, for the potentials considered, the total orbital angular
momentum is a good quantum number \cite{KI,B1}, therefore only the wavefunctions with $L=0$ were included.

It will be shown in the subsequent chapters that both 2-dimensional integral $\mathcal{I}_2$ and 
3-dimensional integral $\mathcal{I}_3$ can be reduced to one-dimensional integrals.
The latter implies that integrals over $\lambda$ and $x$ variables can be evaluated in the explicit (closed)
form.
At that, the final expressions for the integrals associated to 3-dimensional integral (\ref{11})
presents a new results, at least, from mathematical point of view.

\section{Two-dimensional integral}

First, let us consider two-dimensional integral (\ref{10}).
Evaluation of the internal integral over $x$ can be performed by making use of 
the known representation for the associated Laguerre polynomials 
\begin{equation}
\label{20}
\mathcal{L}^{(k)}_{n_1,n_2}(\gamma)\equiv\int_0^\infty L^{(k)}_{n_1}(x)L^{(k)}_{n_2}(x)e^{-\gamma x^2-x}x^k dx=
\sum_{i=0}^{n_1}\sum_{j=0}^{n_2}C_{i,k}^{n_1}~C_{j,k}^{n_2}~\mathcal{J}(k+i+j,\gamma)
\end{equation}
followed by applying the integral presented on p.343(2.3.15(3)) \cite{P1}:
\begin{equation}
\label{21}
\mathcal{J}(\nu,\gamma) \equiv \int_0^\infty e^{-\gamma x^2-x}x^\nu dx=
\Gamma(\nu+1)(2\gamma)^{-\frac{\nu+1}{2}}e^{\frac{1}{8\gamma}}D_{-\nu-1}\left(\frac{1}{\sqrt{2\gamma}} \right). 
\end{equation}
The coefficients introduced in Eq.(\ref{20}) have a form:
\begin{equation}
\label{21a}
C_{i,k}^n=\frac{\left( n-i+1\right)_{k+i}(-1)^i}{(k+i)!i!}.
\end{equation}
Here, $(a)_n$ denotes the Pochhammer symbol, and $D_\sigma(z)$ is the parabolic-cylinder function.
Alternatively, the latter function can be substituted for the Tricomi confluent hypergeometric function
or Hermite function using relations (13.6.36) or (13.6.38), respectively \cite{ABR}:
\begin{equation}
\label{22}
e^{\frac{1}{8\gamma}}D_{-\nu-1}\left(\frac{1}{\sqrt{2\gamma}}\right) =2^{-\frac{\nu+1}{2}}
U \left( \frac{\nu+1}{2},\frac{1}{2},\frac{1}{4\gamma}\right)= 2^{\frac{\nu+1}{2}}
H_{-\nu-1}\left( \frac{1}{2\sqrt{\gamma}}\right). 
\end{equation}
In order to derive the latter connection with Hermite function, we additionally made use
the following relationship (see, e.g., (13.1.29) \cite{ABR})
\begin{equation}
\label{23}
U(a,b,z)=z^{1-b}U(a-b+1,2-b,z),
\end{equation}
that will be applied in the next Section too.

\textit{It is worth noting} that for integer $\nu=n$ the parabolic-cylinder function can be presented in the form \cite{W1}:
\begin{eqnarray}
\label{24}
n!D_{-n-1}(z)=\iu^n 2^{\frac{1-n}{2}} e^{-\frac{z^2}{4}} \sum _{s=1}^n
   (-\iu)^s 
\dbinom{n}{s}
H_{s-1}\left(\frac{z}{\sqrt{2}}\right) H_{n-s}\left(\frac{\iu
   z}{\sqrt{2}}\right) +
~~~~~~~\nonumber~\\
\sqrt{\pi } \iu^n 2^{\frac{1}{2} (-n-1)}
   e^{\frac{z^2}{4}} \text{erfc}\left(\frac{z}{\sqrt{2}}\right)
   H_n\left(\frac{\iu z}{\sqrt{2}}\right),~~~~~~~~~~~~~~~~~
\end{eqnarray}
where $\text{erfc}(y)$ is the complementary error function, $\binom{n}{k}$ is the binomial coefficient,
and $\iu$ represents the imaginary unit.
Eq.(\ref{24}) enables us to derive the additional representation for the integral (\ref{20}):
\begin{equation}
\label{25}
\mathcal{L}^{(k)}_{n_1,n_2}(\gamma)=\frac{(-1)^k}{n_1!n_2!(2\gamma)^{n_1+n_2+k}}
\left[ \frac{1}{2}\sqrt{\frac{\pi}{\gamma}}~e^{\frac{1}{4\gamma}}
\text{erfc}\left( \frac{1}{2\sqrt{\gamma}}\right) T^{(k)}_{n_1,n_2}(\gamma)-S^{(k)}_{n_1,n_2}(\gamma)~ 
\right] 
\end{equation}
where
\begin{eqnarray}
\label{26}
T^{(k)}_{n_1,n_2}(\gamma)=n_1!n_2!(2\gamma)^{n_1+n_2+k}\times
~~~~~~~~~~~~~~~~~~~~~~~~~~~~~~~~~~~~~~~~~~~~~~~~~~~~~~~~~~~~~~\nonumber~\\
\sum_{i=0}^{n_1}
\frac{\left( n_1-i+1\right)_{k+i} }{i!(k+i)!}
\sum_{j=0}^{n_2}\frac{\left( n_2-j+1\right)_{k+j} }{j!(k+j)!}
\left(-\frac{\iu}{2\sqrt{\gamma}} \right)^{i+j+k}
H_{i+j+k}\left(\frac{\iu}{2\sqrt{\gamma}} \right)~
\end{eqnarray}

\begin{eqnarray}
\label{27}
S^{(k)}_{n_1,n_2}(\gamma)=-\frac{n_1!n_2!(2\gamma)^{n_1+n_2+k}}{\sqrt{\gamma}}
\sum_{i=0}^{n_1}
\frac{\left( n_1-i+1\right)_{k+i} }{i!(k+i)!}
\sum_{j=0}^{n_2}\frac{\left( n_2-j+1\right)_{k+j} }{j!(k+j)!}\times
~~~~~~~~~~\nonumber~\\
\left(-\frac{\iu}{2\sqrt{\gamma}} \right)^{i+j+k}~
\sum_{s=1}^{i+j+k}\dbinom{i+j+k}{s}(-\iu)^s
H_{i+j+k-s}\left(\frac{\iu}{2\sqrt{\gamma}} \right)
H_{s-1}\left(\frac{1}{2\sqrt{\gamma}} \right)~~
\end{eqnarray}
\textit{It is worth noting} that both $T^{(k)}_{n_1,n_2}$ and $S^{(k)}_{n_1,n_2}$
present polynomials (in $\gamma$) with the integer coefficients, and possess properties:
\begin{equation}
\label{28}
T^{(k)}_{n_1,n_2}(0)=S^{(k)}_{n_1,n_2}(0)=1.
\end{equation}

\section{Three-dimensional integral}

The aim of this section is to derive the explicit (closed) expression for the internal two-dimensional integral
\begin{equation}
\label{30}
\mathcal{B}^{(k)}_{n_1,n_2}(\beta,\kappa)\equiv
\int^\infty_0 L^{(k)}_{n_1}(x)L^{(k)}_{n_2}(x)e^{-\beta x^2-x} x^k dx
\int_{-1}^1e^{-\lambda \kappa x^2}P_{l_1}(\lambda)P_{l_2}(\lambda)d\lambda
\end{equation}
associated to three-dimensional integral (\ref{11}). To this end, one should first of all to 
apply the well-known Neumann-Adams formula which expresses the product of two Legendre polynomials
as a sum of such polynomials, and then to make use of integral presented by formula (2.17.5(2)), p. 428 \cite{P2}.
This yields:
\begin{eqnarray}
\label{31}
\int_{-1}^1e^{-\lambda \kappa x^2}P_{l_1}(\lambda)P_{l_2}(\lambda)d\lambda=
\sum_{r=0}^{l_1}A^r_{l_1,l_2}\int_{-1}^1 e^{-\lambda \kappa x^2}P_{l_1+l_2-2r}(\lambda) d\lambda=
~~~~~~~~~~\nonumber~\\
x^{-1}\sqrt{\frac{2\pi}{\kappa}}\sum_{r=0}^{l_1} A^r_{l_1,l_2}(-1)^{l_1+l_2-2r}
I_{l_1+l_2-2r+1/2}\left( \kappa x^2\right),~~~~~~~~~~~~(l_1\leq l_2) 
\end{eqnarray}
where
\begin{equation}
\label{32}
 A^r_{l_1,l_2}=\frac{(2l_1-2r-1)!!(2r-1)!!(2l_2-2r-1)!!(l_1+l_2-r)!(2l_1+2l_2-4r+1)}
{(l_1-r)!r!(l_2-r)!(2l_1+2l_2-2r+1)!!},
\end{equation}
and $I_{m+1/2}(z)$ are spherical modified Bessel functions of the first kind ($m$ is integer).

Presenting the product of the Laguerre polynomials in its explicit polynomial form, as it was done
in the previous section, and making use of Eq.(\ref{32}), one obtains:
\begin{eqnarray}
\label{33}
\mathcal{B}^{(k)}_{n_1,n_2}(\beta,\kappa)=\sqrt{\frac{2\pi}{\kappa}}\sum_{r=0}^{l_1} A^r_{l_1,l_2}(-1)^{l_1+l_2-2r}
\sum_{i=0}^{n_1}\sum_{j=0}^{n_2}C_{i,k}^{n_1}~C_{j,k}^{n_2}\times
~~~~~~~~~~~~~~~~~~~~~~~~\nonumber~\\
\int_0^\infty x^{i+j+k-1}e^{-\beta x^2-x}I_{l_1+l_2-2r+1/2}\left( \kappa x^2\right)dx,~~~~~~~~~~
\end{eqnarray}
where coefficients $C_{i,k}^{n}$ are defined by Eq.(\ref{21a}).

Much of what follows are devoted to deriving the explicit analytic expression for evaluating the integral
\begin{equation}
\label{34}
\mathcal{K}_\mu^p(\beta,\kappa)=\int_0^\infty  e^{-\beta x^2-x}I_{\mu}(\kappa x^2)x^pdx.
\end{equation}
We have not found a solution of this problem both in mathematical and physical literature
even for \textit{integer} values of $p$ corresponding to Eq.(\ref{33}).

Final results that will be obtained here are valid for any \textit{real}
$p$ allows the integral (\ref{34}) to be convergent at the given half-integer $\mu$.
On the other hand, it is clear that this integral converges only for $\beta\geq\kappa$.
We shall consider real $\beta>0$ and $\kappa>0$.
Note, that the case of $\beta=\kappa$ is presented in \cite{P2} (see,2.15.6(1), p.306).
However, it is easy to make sure that according to definition (\ref{13}) 
\begin{equation}
\label{35}\
0\leq\frac{\kappa}{\beta}\equiv\frac{\sqrt{3(1-\tau^2)}}{2-\tau}\leq 1
\end{equation}
for values of $-1\leq\tau\leq 1$. 

First, let us present series expansion for the modified Bessel function of the first kind \cite{ABR}:
\begin{equation}
\label{36}\
I_\mu(\kappa x^2)=\left(  \frac{\kappa x^2}{2}\right)^\mu \sum_{k=0}^\infty
\frac{\left(  \frac{\kappa x^2}{2}\right)^{2k}}{ \Gamma(k+\mu+1)k!}.
\end{equation}
Inserting this representation into Eq.(\ref{34}), one obtains:
\begin{eqnarray}
\label{37}
\mathcal{K}_\mu^p(\beta,\kappa)=\left(\frac{\kappa}{2}\right)^{\mu}
\sum_{k=0}^\infty \frac{\left(\frac{\kappa}{2}\right)^{2k}}{k!\Gamma(k+\mu+1)}
\int_0^\infty e^{-\beta x^2-x}x^{4k+2\mu+p}dx=
~~~~~~~~~~~~~~~~~~~~~~~~\nonumber~\\
\frac{\left(\frac{\kappa}{8\beta}\right)^{\mu}}{(4\beta)^{\frac{p+1}{2}}}
\sum_{k=0}^\infty \frac{\Gamma(4k+2\mu+p+1)}{\Gamma(k+\mu+1)k!}
\left(\frac{\kappa}{8\beta}\right)^{2k}U\left(2k+\mu+\frac{p+1}{2},\frac{1}{2}
,\frac{1}{4\beta}\right).~~~~~~
\end{eqnarray}
Instead of Tricomi confluent hypergeometric function, one can use any of equivalent
functions as provided by Eqs.(\ref{21})-(\ref{24}).
\textit{It is worth noting} that the cutoff  expansion (\ref{37}) can be successfully applied for computing integral (\ref{34}) with any values of $(\kappa/\beta)<1$. 
However, the more value of $(\kappa/\beta)$ requires the more length of expansion (\ref{37}). Therefore, the latter is especially effective for very small values of
$(\kappa/\beta)$.

Let us proceed with derivation of the closed analytic expression for the integral (\ref{34}).
The primary definition for the Tricomi confluent hypergeometric function yields \cite{P3} (see, 7.2.2.(2), p.434):
\begin{eqnarray}
\label{38}
U\left(2k+\mu+\frac{p+1}{2},\frac{1}{2}
,\frac{1}{4\beta}\right)=
~~~~~~~~~~~~~~~~~~~~~~~~~~~~~~~~~~~~~~~~~~~~~~~~~~~~~~~~~~~~~~~~~\nonumber~\\
\frac{\sqrt{\pi }\, _1F_1\left(2 k+\frac{p+1}{2}+\mu ;\frac{1}{2};\frac{1}{4
   \beta }\right)}{\Gamma \left(2 k+\frac{p}{2}+\mu +1\right)}-
   \frac{\sqrt{\pi }\, _1F_1\left(2
   k+\frac{p}{2}+\mu +1;\frac{3}{2};\frac{1}{4 \beta }\right)}{\sqrt{\beta }~ \Gamma
   \left(2 k+\frac{p+1}{2}+\mu \right)}.
\end{eqnarray}
Inserting the  first  term of the RHS of Eq.(\ref{38}) into Eq.(\ref{37}) and then changing the order of summation, one obtains:
\begin{eqnarray}
\label{39}
\mathcal{K}_1=\frac{\sqrt{\pi }\left(\frac{\kappa}{8\beta}\right)^{\mu}}{(4\beta)^{\frac{p+1}{2}}}
\sum_{k=0}^\infty \frac{\Gamma(4k+2\mu+p+1)}{\Gamma(k+\mu+1)k!}
\left(\frac{\kappa}{8\beta}\right)^{2k}\frac{\, _1F_1\left(2 k+\frac{p+1}{2}+\mu ;\frac{1}{2};\frac{1}{4
   \beta }\right)}{\Gamma \left(2 k+\frac{p}{2}+\mu +1\right)}=
~~~~~\nonumber~\\
\frac{\sqrt{\pi} \left( \frac{\kappa}{2\beta}\right)^\mu }{2\beta^\frac{p+1}{2}}
\sum_{n=0}^\infty \frac{1}{n!\Gamma(1/2+n)(4\beta)^n}\sum_{k=0}^\infty \frac{\Gamma(2k+\frac{p+1}{2}+\mu+n)}
{k!\varGamma(k+\mu+1)}\left( \frac{\kappa}{2\beta}\right)^{2k}=
~~~~~~~~~~\nonumber~\\
\frac{\left(\frac{\kappa}{2\beta}\right)^{\mu}}{2\beta^\frac{p+1}{2}}
\sum_{n=0}^\infty \frac{\Gamma\left( n+\mu+\frac{p+1}{2}\right) }{(2n)!\beta^n}
\,_2F_1\left( \frac{n+\mu}{2}+\frac{p+1}{4},\frac{n+\mu}{2}+\frac{p+3}{4};1+\mu;\frac{\kappa^2}{\beta^2}\right),~~~~~~~~ 
\end{eqnarray}

We used for the latter derivation and we will use in what follows, the well-known duplication
formula (6.1.18) \cite{ABR} for the gamma functions.

Next step is making use of the relationship between the Gauss hypergeometric functions 
of the form $\,_2F_1( a,a+1/2;c;z^2)$
and the associated Legendre functions of the first kind \cite{P3} (7.3.1(101), p.460), which yields for the case of interest:
\begin{eqnarray}
\label{40}
\,_2F_1\left( \frac{n+\mu}{2}+\frac{p+1}{4},\frac{n+\mu}{2}+\frac{p+3}{4};1+\mu;z^2\right)=
~~~~~~~~~~~~~~~~~~~~~~~~~~~~~~~\nonumber~\\
\Gamma(\mu+1)\left( \frac{2}{\iu z}\right)^\mu (1-z^2)^{-\frac{2n+p+1}{4}}
P_{n+\frac{p-1}{2}}^{-\mu}\left( \frac{1}{\sqrt{1-z^2}}\right),~~~~~  
\end{eqnarray}
with $z=\kappa/\beta$.

In this stage, one needs to express the Legendre functions presented in Eq.(\ref{40}) through finite sums
with the limits (of summation), which are not dependent on $n$.
Note, that relationship 3.2(9) \cite{BE}  enables us to express the associated Legendre functions of
the first kind through two Gauss hypergeometric functions by different manners, which are 
presented by formulas (3.2.14)-(3.2.31) \cite{BE}. From the latter representations it is seen that only 
the case of a half-integer $\mu$ could give the mentioned above Gauss hypergeometric functions with one
(of the first two) negative integer parameter, which is not dependent on $n$.

Thus, introducing denotation\begin{equation}
\label{41}
\mu=m+\frac{1}{2}, 
\end{equation}
where $m$ is a non-negative integer, and making use representation 3.2(30) \cite{BE}, one obtains:
\begin{eqnarray}
\label{42}
P_{n+\frac{p-1}{2}}^{-m-\frac{1}{2}}\left( \frac{1}{\sqrt{1-z^2}}\right)=
\frac{e^{\iu \pi\left( \frac{m}{2}-\frac{3}{4}\right) }}{\sqrt{2 \pi z}}\times
~~~~~~~~~~~~~~~~~~~~~~~~~~~~~~~\nonumber~\\
\left[ \sqrt{1-z}\left(\frac{1-z}{1+z} \right)^{\frac{2n+p-1}{4}} \frac{(-1)^m \Gamma(n+p/2)}{\Gamma(n+p/2+m+1)}
\,_2F_1\left( m+1,-m;1-n-\frac{p}{2};\frac{1+z}{2z}\right)-
\right. 
~~\nonumber~\\
\left. 
\sqrt{1+z}\left(\frac{1+z}{1-z} \right)^{\frac{2n+p-1}{4}} \frac{ \Gamma(n+p/2-m)}{\Gamma(n+p/2+1)}
\,_2F_1\left( m+1,-m;1+n+\frac{p}{2};\frac{1+z}{2z}\right) 
\right].~~~~~~~ 
\end{eqnarray}
Note, that factor $e^{\iu \pi\left( \frac{m}{2}-\frac{3}{4}\right)}$ is correct, but it is different from the corresponding factor presented in Ref.\cite{BE}.

One should emphasize, that due to the factor $\Gamma(n+p/2-m)$ 
Eq.(\ref{42}) is not valid only for the case of even $p\leq 2m$ $(n\geq0)$. This case will be considered later.

Now, one needs to insert Eq.(\ref{42}) with the hypergeometric functions presented in explicit (polynomial) form
into Eq.(\ref{40}). Then, inserting the result into  Eq.(\ref{39}) and changing the order of summation,
one obtains for the real values of $p>-2m-2$,  excluding the case of even $p\leq2m$:
\begin{eqnarray}
\label{43}
\sqrt{8 \pi \kappa}~ \mathcal{K}_1\left(\neg~ \text{even}~ p\leq 2m \right)=
~~~~~~~~~~~~~~~~~~~~~~~~~~~~~~~~~~~~~~~~~~~~~~~~~~~~~~~~~~~~~~~~~~~~\nonumber~\\
\left[ 
\left( \frac{1}{\beta-\kappa}\right) ^{p/2}\sum_{n=0}^\infty \frac{\Gamma\left(m+n+1+\frac{p}{2} \right)\Gamma\left(n+\frac{p}{2}-m\right)  }{(2n)!\Gamma\left( 1+n+\frac{p}{2}\right)(\beta-\kappa)^n}
\,_2F_1\left( m+1,-m;1+n+\frac{p}{2};\frac{\beta+\kappa}{2\kappa}\right)-
\right.  
~\nonumber~\\
\left. 
\left( \frac{1}{\beta+\kappa}\right) ^{p/2}(-1)^m\sum_{n=0}^\infty \frac{\Gamma\left(n+\frac{p}{2} \right) }
{(2n)!(\beta+\kappa)^n}
\,_2F_1\left( m+1,-m;1-n-\frac{p}{2};\frac{\beta+\kappa}{2\kappa}\right)
\right]= 
~\nonumber~\\
(\beta-\kappa)^{-\frac{p}{2}}\Gamma\left(\frac{p}{2}+m+1\right)\Gamma\left(\frac{p}{2}-m\right)\times
~~~~~~~~~~~~~~~~~~~~~~~~~~~~~~~~~~~~~~~~~~~~~~~~~~~~~~~~~~~~~~~~~~\nonumber~\\
\sum_{k=0}^m\frac{(m+k)!\left( -\frac{\beta+\kappa}{2\kappa}\right)^k }{k!(m-k)!\Gamma\left(1+k+\frac{p}{2} \right) }
\,_2F_2\left[\frac{p}{2}-m,\frac{p}{2}+m+1;\frac{1}{2},\frac{p}{2}+k+1;\frac{1}{4(\beta-\kappa)}\right]-
~~~~~~~~\nonumber~\\
( \beta+\kappa)^{-\frac{p}{2}}(-1)^m
\sum_{k=0}^m \frac{(m+k)!\Gamma\left( \frac{p}{2}-k \right)\left(\frac{\beta+\kappa}{2\kappa} \right)^k }
{k!(m-k)!}
\,_1F_1\left[\frac{p}{2}-k;\frac{1}{2};\frac{1}{4(\beta+\kappa)}\right].~~~~~~~~~~~~~~~~~~~~~~~
  \end{eqnarray}
For the even values of $p\leq 2m$, one needs to divide a summation over $n$ by two ranges:
$[0:m-p/2]$ and  $[m-p/2+1:\infty]$, and then - to perform the same procedure, as in 
the previous case. This yields:
\begin{eqnarray}
\label{44}
\sqrt{8 \pi \kappa}~ \mathcal{K}_1\left(\text{even}~ p\leq 2m \right)=
~~~~~~~~~~~~~~~~~~~~~~~~~~~~~~~~~~~~~~\nonumber~\\
\frac{2\sqrt{\pi}\left(\frac{\kappa}{2\beta} \right)^{m+1} }{\Gamma\left(m+\frac{3}{2} \right) \beta^\frac{p}{2}}
\sum_{n=0}^{m-\frac{p}{2}}\frac{\Gamma\left(\frac{p}{2}+m+n+1 \right) }{(2n)!\beta^n}
\,_2F_1\left( \frac{n+m+1}{2}+\frac{p}{4},\frac{n+m+2}{2}+\frac{p}{4};m+\frac{3}{2};\frac{\kappa^2}{\beta^2}\right)+
~~~~\nonumber~\\
\frac{1}{(2m+2-p)!}\left\lbrace
\left(- \frac{1}{\beta+\kappa}\right)^{m+1}\times
\right. 
~~~~~~~~~~~~~~~~~~~~~~~\nonumber~\\
\left. 
\sum_{k=0}^m \frac{(m+k)!\left( \frac{\beta+\kappa}{2\kappa}\right)^k }{k!}
\,_2F_2\left[1,1-k+m;m+\frac{3-p}{2},m+2-\frac{p}2{};\frac{1}{4(\beta+\kappa)}\right]+
\right. 
~~\nonumber~\\ 
\left(\frac{1}{\beta-\kappa}\right)^{m+1}(2m+1)!\times
~~~~~~~~~~~~~~~~~~~~~~\nonumber~\\ 
\left. 
\sum_{k=0}^m \frac{\left( -\frac{\beta+\kappa}{2\kappa}\right)^k }{k!(m-k)!(m+k+1)}
\,_3F_3\left[1,1,2+2m;2+k+m,m+\frac{3-p}{2},2+m-\frac{p}{2};\frac{1}{4(\beta-\kappa)}\right]
\right\rbrace. 
~~\nonumber~\\ 
\end{eqnarray}
One can proceed to consideration of the residual RHS of Eq.(\ref{37}).
Inserting the second term of the RHS of Eq.(\ref{38}) into Eq.(\ref{37}) and then changing the order of summation, one obtains:
\begin{eqnarray}
\label{45}
\mathcal{K}_2=-\frac{\sqrt{\pi }\left(\frac{\kappa}{8\beta}\right)^{m+1/2}}{(4\beta)^{\frac{p+1}{2}}}
\sum_{k=0}^\infty \frac{\Gamma(4k+2m+p+2)}{k!\Gamma(k+m+3/2)}
\left(\frac{\kappa}{8\beta}\right)^{2k}\frac{\, _1F_1\left(2 k+\frac{p+3}{2}+m;\frac{3}{2};\frac{1}{4
   \beta }\right)}{\sqrt{\beta}~\Gamma \left(2 k+\frac{p}{2}+m+1\right)}=
~~~~~~~~~~\nonumber~\\
-\frac{\left(\frac{\kappa}{2\beta}\right)^{m+1/2}}{2\Gamma(m+3/2)\beta^{\frac{p}{2}+1}}
\sum_{n=0}^\infty \frac{\Gamma\left(n+m+\frac{p+3}{2} \right) }{(2n+1)!\beta^n}
\,_2F_1\left[\frac{n+m}{2}+\frac{p+3}{4},\frac{n+m}{2}+\frac{p+5}{4};m+\frac{3}{2};\frac{\kappa^2}{\beta^2}\right].~~~
\end{eqnarray}
Once again one should make use of the relationship between the Gauss hypergeometric functions 
of the form $\,_2F_1( a,a+1/2;c;z^2)$
and the associated Legendre functions of the first kind \cite{P3} (7.3.1(101), p.460). However, in  this case
one sets $a=(n+m)/2+(p+3)/4,~c=m+3/2,~z=\kappa/\beta$. Then, one needs to apply representation 3.2(9) 
 \cite{BE} with the parameters presented by 3.2(30) \cite{BE}. Thus, one obtains:
\begin{eqnarray}
\label{46}
\,_2F_1\left[\frac{n+m}{2}+\frac{p+3}{4},\frac{n+m}{2}+\frac{p+5}{4};m+\frac{3}{2};z^2\right]=
\frac{\Gamma(m+\frac{3}{2})}{2\sqrt{\pi(1-z^2)}}\left(\frac{2}{z} \right)^{m+1}\left( 1-z^2\right)^{-\frac{2n+p}{4}} 
~~\nonumber~\\
\left[ 
\sqrt{1+z}\left(\frac{1+z}{1-z}\right)^{\frac{2n+p}{4}}
\frac{\Gamma\left(n+\frac{p+1}{2}-m\right) }{\Gamma\left(n+\frac{p+3}{2}\right)} 
\,_2F_1\left(m+1,-m;n+\frac{p+3}{2};\frac{z+1}{2z}\right)-
\right. 
~~~~\nonumber~\\
\left. 
\sqrt{1-z}\left(\frac{1-z}{1+z}\right)^{\frac{2n+p}{4}}
\frac{(-1)^m\Gamma\left(n+\frac{p+1}{2}\right) }{\Gamma\left(n+\frac{p+3}{2}+m\right)}
\,_2F_1\left(m+1,-m;-n+\frac{1-p}{2};\frac{z+1}{2z}\right)
\right].~~~~~~ 
\end{eqnarray}
One should notice that due to the factor $\Gamma\left(n+\frac{p+1}{2}-m\right)$ Eq.(\ref{46})
is not valid for odd values of $p\leq2m-1$.

Substituting representation (\ref{46}) with the explicit (polynomial) expressions for the hypergeometric functions
into Eq.(\ref{45}), and then changing the order of summation, one obtains:
\begin{eqnarray}
\label{47}
\sqrt{8 \pi \kappa}~ \mathcal{K}_2\left(\neg~ \text{odd}~ p\leq 2m-1 \right)=
~~~~~~~~~~~~~~~~~~~~~~~~~~~~~~~~~~~~~~~~\nonumber~\\
(-1)^m (\beta+\kappa)^{-\frac{p+1}{2}}\sum_{k=0}^m \frac{(m+k)!\Gamma\left(\frac{p+1}{2}-k \right) }
{(m-k)!k!} \left( \frac{\beta+\kappa}{2\kappa}\right)^k 
\,_1F_1\left[\frac{p+1}{2}-k;\frac{3}{2};\frac{1}{4(\beta+\kappa)}\right]-
~~~~\nonumber~\\
 (\beta-\kappa)^{-\frac{p+1}{2}}\Gamma\left(\frac{p+1}{2}-m\right)\Gamma\left(\frac{p+3}{2}+m\right)\times
~~~~~~~~~~~~~~~~~\nonumber~\\ 
\sum_{k=0}^m \frac{(m+k)!(-1)^k}{(m-k)!k!\Gamma\left( \frac{p+3}{2}+k\right) }
\left( \frac{\beta+\kappa}{2\kappa}\right)^k 
\,_2F_2\left[\frac{p+1}{2}-m,\frac{p+3}{2}+m;\frac{3}{2},\frac{p+3}{2}+k;\frac{1}{4(\beta-\kappa)}\right].
~~~~~~~~ 
\end{eqnarray}
For the odd values of $p\leq 2m-1$, one needs to divide a summation over $n$ in Eq.(\ref{45}) by two ranges:
$[0:m-(p+1)/2]$ and  $[m-(p-1)/2:\infty]$. And then, it is necessary to perform the same procedure, as in 
the previous case. This yields:
\begin{eqnarray}
\label{48}
\sqrt{8 \pi \kappa}~ \mathcal{K}_2\left(\text{odd}~ p\leq 2m-1 \right)=
~~~~~~~~~~~~~~~~~~~~~~~~~~~~~~\nonumber~~~~\\
-\frac{2\sqrt{\pi}\left(\frac{\kappa}{2\beta} \right)^{m+1} }{\Gamma\left(m+\frac{3}{2} \right) \beta^\frac{p+1}{2}}
\sum_{n=0}^{m-\frac{p+1}{2}}\frac{\Gamma\left(m+n+\frac{p+3}{2} \right) }{(2n+1)!\beta^n}
\,_2F_1\left( \frac{n+m}{2}+\frac{p+3}{4},\frac{n+m}{2}+\frac{p+5}{4};m+\frac{3}{2};\frac{\kappa^2}{\beta^2}\right)-
~~\nonumber~~~~\\
\frac{1}{(2m+2-p)!}\left\lbrace
\left(- \frac{1}{\beta+\kappa}\right)^{m+1}\times
\right. 
~~~~~~~~~\nonumber~~~~~~\\
\left. 
\sum_{k=0}^m \frac{(m+k)!\left( \frac{\beta+\kappa}{2\kappa}\right)^k }{k!}
\,_2F_2\left[1,1-k+m;m+\frac{3-p}{2},m+2-\frac{p}2{};\frac{1}{4(\beta+\kappa)}\right]+
\right. 
~~~\nonumber~~~~~~~~\\ 
\left(\frac{1}{\beta-\kappa}\right)^{m+1}(2m+1)!\times
~~~~~\nonumber~~~~~~~~~~\\ 
\left. 
\sum_{k=0}^m \frac{\left( -\frac{\beta+\kappa}{2\kappa}\right)^k }{k!(m-k)!(m+k+1)}
\,_3F_3\left[1,1,2+2m;2+k+m,m+\frac{3-p}{2},2+m-\frac{p}{2};\frac{1}{4(\beta-\kappa)}\right]
\right\rbrace.~~~~~ 
\end{eqnarray} 

\subsection{Final formula for the new integral}

The results obtained above for the integral (\ref{34}) with half-integer parameter $\mu$
can be processed and presented in the compact form. To this end, let us introduce 
three auxiliary functions:
\begin{eqnarray}
\label{51}
\mathcal{F}_1(s)\equiv\mathcal{F}_1(s,p,m;\beta,\kappa)=
\frac{2\sqrt{\pi}\left(\frac{\kappa}{2\beta} \right)^{m+1} }{\Gamma\left(m+\frac{3}{2} \right) \beta^\frac{p+s}{2}}\times
~~~~~~~~~~~~~~~~~~~~~~~~\nonumber~\\
\sum_{n=0}^{m-\frac{p+s}{2}}\frac{\Gamma\left(m+n+1+\frac{p+s}{2} \right) }{(2n+s)!\beta^n}
\,_2F_1\left( \frac{n+m+1}{2}+\frac{p+s}{4},\frac{n+m+2}{2}+\frac{p+s}{4};m+\frac{3}{2};\frac{\kappa^2}{\beta^2}\right),~~~~
\end{eqnarray}
\begin{eqnarray}
\label{52}
\mathcal{F}_2\equiv \mathcal{F}_2(p,m;\beta,\kappa)=
\frac{1}{(2m+2-p)!}\left\lbrace
\left(- \frac{1}{\beta+\kappa}\right)^{m+1}\times
\right. 
~~~~~~~~~~\nonumber~~~~~~~~~~~~~~~\\
\left. 
\sum_{k=0}^m \frac{(m+k)!\left( \frac{\beta+\kappa}{2\kappa}\right)^k }{k!}
\,_2F_2\left[1,1-k+m;m+\frac{3-p}{2},m+2-\frac{p}2{};\frac{1}{4(\beta+\kappa)}\right]+
\right. 
\frac{(2m+1)!}{(\beta-\kappa)^{m+1}}\times
~~\nonumber~~~~\\ 
\left. 
\sum_{k=0}^m \frac{\left( -\frac{\beta+\kappa}{2\kappa}\right)^k }{k!(m-k)!(m+k+1)}
\,_3F_3\left[1,1,2+2m;2+k+m,\frac{3-p}{2}+m,2+m-\frac{p}{2};\frac{1}{4(\beta-\kappa)}\right]
\right\rbrace,~~~~ 
\end{eqnarray}
\begin{eqnarray}
\label{53}
\mathcal{F}_3(s)\equiv\mathcal{F}_3(s,p,m;\beta,\kappa)=
~~~~~~~~~~~~~~~~~~~~~~~~~~~~~~~~~~~~~~\nonumber~~~~~~~~\\
\frac{(-1)^m}{(\beta+\kappa)^{\frac{p+s}{2}}}
\sum_{k=0}^m \frac{(m+k)!\Gamma\left(\frac{p+s}{2}-k \right) }
{(m-k)!k!} \left( \frac{\beta+\kappa}{2\kappa}\right)^k 
\,_1F_1\left[\frac{p+s}{2}-k;\frac{2s+1}{2};\frac{1}{4(\beta+\kappa)}\right]+
~~~~~~~~~~\nonumber~~~~\\
\frac{\Gamma\left(\frac{p+s}{2}-m\right)\Gamma\left(\frac{p+s}{2}+m+1\right)}
{(\beta-\kappa)^{\frac{p+s}{2}}} 
\sum_{k=0}^m \frac{(m+k)!(-1)^k}{(m-k)!k!\Gamma\left( \frac{p+s}{2}+k+1\right) }
\left( \frac{\beta+\kappa}{2\kappa}\right)^k \times
~~~~~~~~~~~~~~~~\nonumber~~~~~~~\\
\,_2F_2\left[\frac{p+s}{2}-m,\frac{p+s}{2}+m+1;\frac{2s+1}{2},\frac{p+s}{2}+k+1;\frac{1}{4(\beta-\kappa)}\right].
~~~~~~~~~~~~~
\end{eqnarray}
The integral of interest can be expressed in term of these functions as follows:
\begin{equation}
\label{54}
\int_0^\infty e^{-\beta x^2-x}I_{m+\frac{1}{2}}(\kappa x^2)x^p dx=\frac{1}{\sqrt{8 \pi \kappa}}
\begin{cases}
~~\mathcal{F}_1(0)+ \mathcal{F}_2-\mathcal{F}_3(1)~~~~~~~~\text{even}~p\leq2m \\
-\mathcal{F}_1(1)- \mathcal{F}_2+\mathcal{F}_3(0)~~~~~~~~\text{odd}~p< 2m \\
~~\mathcal{F}_3(0)-\mathcal{F}_3(1)~~~~~~~~~~~~~~~~~\text{otherwise}
\end{cases}
\end{equation}
It is seen that representation (\ref{54}) 
cannot be used for very small values of parameter $\kappa$.
However, in this case, the cutoff representation (\ref{37}) can be applied with advantage.
Moreover, one should notice that the closed form (\ref{54}) presents the difference of two large quantities,
because $\mathcal{K}_1$ and $\mathcal{K}_2$ have different signs, but their absolute values are very close.
At that, the relation $|\mathcal{K}_1/(\mathcal{K}_1-\mathcal{K}_2)|$   
increases very quickly with parameter $m$ of the spherical modified Bessel
function of the first kind. Therefore,  the quantities  $\mathcal{K}_1$ and $\mathcal{K}_2$ 
(i.e., functions $\mathcal{F}_i$)
have to be calculated with high accuracy, especially  for large $m$.

\textit{It is worth noting} that a few-body problem is related to the integral (\ref{54}) with integer
power $p$ and even $m$, whereas formulas (\ref{51})-(\ref{54}) are valid for any integer $m\geq0$ and any
real $p>-2m-2$.

\newpage

\end{document}